\documentstyle[12pt,a4]{article}
\renewcommand{\baselinestretch}{1.6}
\begin{document}
\Large
\begin{center}
\bf{Ab initio simulations of liquid systems:
\\Concentration dependence
of the electric conductivity of NaSn alloys}\\[2cm]
\normalsize
\em
R. Kaschner, M. Sch\"one, and G. Seifert\\
Technische Universit\"at Dresden\\ Institut f\"ur
Theoretische Physik\\
D-01062 Dresden, Germany\\[.3cm]
G. Pastore\\ Istituto Nazionale di Fisica della Materia
and \\
Dipartimento di Fisica Teorica dell' Universit\`{a}\\
Strada Costiera 11\\I-34014 Trieste, Italy\\[.7cm]
\em
\end{center}
\normalsize

Liquid NaSn alloys in five different
compositions (20, 40, 50, 57 and 80\% sodium) are
studied using density
functional calculations combined with molecular dynamics
(Car-Parrinello method). The frequency-dependent
electric conductivities for
the systems
are calculated by means of the Kubo-Greenwood formula.\\
The extrapolated
DC conductivities are in good agreement with the
experimental data and
reproduce the strong variation with the concentration.
The maximum of conductivity is obtained, in agreement
with experiment, near
the equimolar composition.\\
The strong variation of conductivity, ranging from
almost semiconducting up to
metallic behaviour, can be understood by an analysis of
the densities-of-states.

\newpage
\renewcommand{\baselinestretch}{1.0}
\large
\normalsize
During the past decades much work has been performed to
investigate
binary alloys, among which the so-called Zintl systems
are of special interest.
One typical example of these Zintl systems are
alloys made of alkali metals and elements of the fourth
group of the periodic table.
In the last years much effort was made to investigate
these systems
in the liquid phase not only experimentally but also
theoretically.
One of the most successful theoretical tools to describe
and interpret
experimental findings is to perform \em ab initio \em
molecular dynamics
(MD) simulations for the particular systems.
{}From the obtained trajectories one can extract data such
as different structural and electronic properties of the
liquid
alloys. These results can be compared directly with
experimental data.

Detailed investigations have already been performed
for some Zintl alloys, such as K-Si \cite{KSi},
equimolar NaSn \cite{SPCar},
Li-Si \cite{LiSi} and Cs-Pb \cite{CsPb}.
However, these investigations were limited
to only one composition.
In a recent paper \cite{Schoene} we extended the
investigation of the liquid NaSn alloys to a wide
range of compositions ranging from 20\% up to 80\% of
sodium.
\em Ab-initio \em MD allowed to simulate the change of
structural properties in these systems and made possible
to discuss
the static structure factors and the behaviour of the
Zintl anions (Sn$_4^{4-}$) in the liquid phase.
\cite{Schoene}.

After the analysis of the structural properties, we have
undertaken a detailed study of the electronic properties
of these alloys and in the present letter we report the
main results of such an analysis.
In particular, we have calculated the electric
conductivity as function of
composition by averaging the Kubo-Greenwood formula
\cite{KG} over the MD trajectories obtained in
\cite{Schoene}.

In the present study for the first time, we analyze,
by means of computer simulation,
the strong variation of the conductivity - ranging from
metallic to almost
semiconducting behaviour - with concentration.
To get more insight into the electronic properties, we
evaluated the electronic density of states and analyzed
its specific atomic
contributions.

We will start with the description of the methodology.
After that, the results are presented. Finally a summary
and outlook is
given.\\~\\

The MD simulations discussed in this paper were
performed using the
Car-Parrinello method \cite{CP} applying the MOTECC-90
computer code \cite{Hohl}. \\
The considered systems consist of 64 atoms in a cubic
unit cell with a length
of 23.4 a.u. and periodic boundary conditions.
The plane-wave cut-off was chosen to be 6 Ryd which we
have justified
by tests for dimers and bulk systems \cite{Schoene}.
We used the pseudopotentials of Bachelet et al.
\cite{BHS}.\\
The data for our analysis were collected from
``production  runs'' of approximately
10000 steps each, corresponding to a total simulation
time of about 2 ps. The
temperature for each simulation was about 50 K above
the experimental liquidus curve in the phase diagram.\\
For a more detailed description of computational
features and the simulation procedure (including systems
and
temperatures) see \cite{Schoene,bigpaper}.\\

The electric (AC) conductivity for a single
configuration of
the considered systems is obtained
using the Kubo-Greenwood formula \cite{KG}:

$$\sigma(\omega)=\frac{2\pi e^2}{3m^2\omega \Omega}
\sum_m^{occ}\sum_n^{unocc}
 \sum_{\alpha}
|\langle\psi_m|\hat{p}_{\alpha}|\psi_n)\rangle|^2 \cdot
\delta(\varepsilon_n
 -\varepsilon_m-\hbar\omega),$$

where $m$ and $e$ are the electronic mass and charge,
respectively. $\Omega$
is the MD cell volume and $\hat{p}_{\alpha}$ is the
$\alpha$ component of the momentum operator.
The sum over $m$ and $n$ are respectively over the
occupied and unoccupied
states corresponding to the one-particle eigenvalues
$\varepsilon_m$ and $\varepsilon_n$.
The extrapolation $\omega \to 0$ gives the DC
conductivity.

In principle, to evaluate the thermal average of
$\sigma(\omega)$,
one should average the values obtained
for all the configurations of an MD trajectory.
However, this is not necessary, because
consecutive time steps correspond to highly
correlated configurations. Therefore, a significant
saving of
computational time can be gained
if the configurations used to perform the average are
spaced (well ``separated'') in time.
The separation between each selected configuration
should be large to ensure
statistical independence between the configurational
contributions to the
average. We have chosen to perform a more drastic
reduction
of the contributions
to the average due to the very heavy demand of the
calculation of $\sigma(\omega)$
- taking in mind that the need of
computing time is high since the evaluation of the
Kohn-Sham states
required by the Kubo-Greenwood formula was performed for
a large number of states (200).
In particular, we chosed a time interval between
configurations such
that the mean squared displacement of the ions is
about the size of the nearest-neighbour distance.
For our cases it was sufficient to take into account
$\sigma(\omega)$ for 6 geometries for each composition.
We think this number of configurations should be
considered as reasonable for a first analysis.
Future work should exploit the possibility of performing
calculations for a small set of empty states {\it on the
fly} when the MD simulation runs. This type of
calculations
should be possible also within LCAO schemes.\\

To interpret the strong dependence of the conductivity
from composition we calculated the averaged
density-of-states (DOS)
considering the same configurations as described above
for each case.
The electronic DOS has been evaluated using the same
description
of the electronic states as in the Car-Parrinello (plane
wave expansion) MD calculations and also in a more
approximate way by using a
simplified LCAO-DFT method \cite{LCAO}. However, at
variance with the plane wave
expansion, the LCAO method allows to split in a rather
unambiguous way the total DOS
into fractions (i.e. partial densities-of-states)
referring to the sodium and tin atoms,
respectively.\\~\\

The AC conductivity $\sigma(\omega)$ obtained for one
representative composition is given in Fig.~1.
The obtained curves $\sigma(\omega)$ indicate that
the frequency dependence is non Drude-like
witnessing the non-free
electron like nature of the electronic states.
For the extrapolation towards $\omega \to 0$ the data
for small $\omega$ must be handled carefully.
As in this region the denominator in the Kubo-Greenwood
formula becomes very
small and the statistical noise is very high, the error
in the low frequency part is considerably large
(cf. the discussion in \cite{LiSi}).
In Fig.~1 the extrapolated value of $\sigma(0) =
\sigma_{DC}$ is marked by a triangle.
The statistical error of the calculated resistivity
$\rho = 1/\sigma(0)$ is for each case about $\pm$100
$\mu\Omega$cm.

Fig. 2 shows our calculated resistivities in comparison
with the
experimental figures by van der Marel et al. \cite{Exp},
which have been measured for 70 K above the liquidus.
As can be seen, the agreement is, within the statistical
error,
very good. The only exception is the composition with 80
\% tin;
however, the trend to a small (metallic)
resistivity is obtained correctly for this
case. For the discussion of this agreement it should be
noted that
one has also (i) approximations
underlying the usage of the Kubo-Greenwood formula
generally
\cite{KG} and (ii) small
differences between the experimental and theoretical
temperatures (see above) and densities \cite{Schoene}.

The most important result is
the faithful reproduction of the variation of the
resistivity with the composition shown by the
experimental data.

This strong dependence of the conductivity on the
composition can be understood qualitatively
considering the densities of states. This discussion
will be given in detail in a forthcoming paper
\cite{bigpaper}.
Here we only summarize the essential results:

The DOS splits up into contributions from
the components (Na, Sn) of the alloy. Both contributions
are well
separated in energy (due to the atomic Kohn-Sham
energies of sodium and tin):
The low-energy region is dominated by tin, whereas for
high energies
sodium becomes dominant.

For compositions with a small content
of sodium the Fermi energy lies in the region dominated
by tin. This corresponds to a high DOS and therefore to
a high (metallic) conductivity.

For compositions with 50 and 57\% sodium the Fermi level
is out of the
tin-dominated region, but still has not reached the
sodium-dominated area.
This yields a small DOS for these cases at the Fermi
level.
Therefore, one gets an explanation for the minimum of
the conductivity near the
equimolar composition, as can be seen in Fig. 2.
Analogously,
for solid equimolar $\beta$-NaSn even a gap at the Fermi
level (between two sets
of states dominated by tin and sodium, respectively) was
reported in literature \cite{Gap}.

In compositions with excess sodium the situation changes
again: The
Fermi level is in the sodium-dominated region, yielding
again a metallic
behaviour - similar to the tin-rich cases.

The observed behavior of the DOS is consistent with the
theoretical analysis proposed by Geertsma \cite{Geertsma}
some years ago for these systems. A more detailed
comparison between our
simulation data and Geertsma's analysis is in progress.\\~\\
We have determined the electric conductivities of liquid
Na-Sn alloys
for five different compositions with the Kubo-Greenwood
scheme,
using the trajectories from our \em ab initio \em MD
simulations.
The calculated values reproduce the measured strong
variation of the
conductivity with the Na (or Sn) concentration very
well. The
semiconductor-like conductivity of the alloys with
40...60\% Na can
be understood by considering the behaviour of the
densities-of-states.

Our calculations were performed for temperatures about
50 K above
the liquidus. However, the experiments \cite{Exp}
yielded also
a considerable temperature dependence of the
conductivities in the liquid and in the
solid phases \cite{Sab}. Hence, it is desirable to
consider also
(i) the $T$ dependence of the conductivity for the
liquid alloys, (ii)
the conductivities of the solid phases, in particular,
for the equimolar case
\cite{Sab}. Such investigations will be a subject of our
future work.

In addition, we are going to consider larger supercells
to investigate
the effects of the finite supercell using the
above-mentioned LCAO scheme \cite{LCAO}.\\

{\large \bf Acknowledgements}\\~\\
Parts of the calculations have been performed on the
Cray systems  of the Centro
Interdipartimentale di Calcolo dell'
Universit\`a di Trieste and of CINECA at
Bologna. G.P. acknowledges the CNR for the financial support
Contributo di ricerca n. 95.00505.CT12 .
This work was supported by the Deutsche
Forschungsgemeinschaft (DFG).

\vspace*{1cm}

{\Large\bf Figure captions}\\~\\

Figure 1: $\sigma(\omega)$: AC conductivity
$\sigma(\omega)$
for the composition with 50\% tin
calculated with the Kubo-Greenwood formula. The plot
represents an average over six configurations taken from
\em ab initio \em MD simulations.\\
The extrapolated value for the DC conductivity is
indicated by a triangle.\\

\vspace*{0.5cm}

Figure 2: Calculated resistivities $\rho=1/\sigma(0)$
(dots) with error bars (resulting from the extrapolation and
from the configuration average)
compared with experimental values (solid line)
by van der Marel et al. \cite{Exp} referring to
systems 70 K above the liquidus.\\


\begin{thebibliography}{99}

\bibitem{KSi} G. Galli and M. Parrinello, J. Chem. Phys.
 {\bf 95}, 7054 (1991).
\bibitem{SPCar} G. Seifert, G. Pastore, and R. Car, J.
 Phys. Condensed Matter {\bf 4}, L179 (1992).
\bibitem{LiSi} G.A. de Wijs, G. Pastore, A. Selloni, and
 W. van der Lugt, Phys. Rev. B {\bf 48}, 13459 (1993).
\bibitem{CsPb} G.A. de Wijs, G. Pastore, A. Selloni, and
 W. van der Lugt, Europhys. Lett. {\bf 27}, 667 (1994).
\bibitem{Schoene} M. Sch\"one, R. Kaschner, and G.
 Seifert, J. Phys. Condensed Matter {\bf 7}, L19 (1995).
\bibitem{KG} E.N. Economou, \em Green's Functions in
 Quantum Physics\em,
 Springer Series in Solid State Sciences, Vol. 7,
 Springer, Berlin, 1983, p.152 ff; for a general discussion
 about the usage of the Kubo-Greenwood formula for liquids
 within density functional theory see also F. Kirchhoff, J.M. Holender,
 and M.J. Gillan, Phys. Rev. B {\bf 54}, 190 (1996).
\bibitem{CP} R. Car and M. Parrinello, Phys. Rev. Lett.
 {\bf 55}, 2471 (1985).
\bibitem{Hohl} D. Hohl, Density Functional Theory
 Workshop, Cornell Theory Center, 1992.
\bibitem{BHS} G.B. Bachelet, D.R. Hamann, and M.
 Schl\"uter, Phys. Rev. B {\bf 62}, 4199 (1982);
 D.R. Hamann, Phys. Rev. B {\bf 40}, 2980 (1989).
\bibitem{bigpaper} G. Seifert, G. Pastore, R. Kaschner,
 and M. Sch\"one, to be published.
\bibitem{LCAO} D. Porezag, Th. Frauenheim, Th. K\"ohler,
 G. Seifert, and R. Kaschner, Phys. Rev. B {\bf 51},
 12947 (1995);
 G. Seifert, D. Porezag, and Th. Frauenheim, Int. J.
 Quantum Chem.
 {\bf 58}, 185 (1996).
\bibitem{Exp} C. van der Marel, A.B. van Oosten, W.
 Geertsma, and W. van der Lugt, J. Phys. F {\bf 12},
2349 (1982).
\bibitem{Gap} F. Springelkamp, R.A. de Groot, W.
 Geertsma, W. van der Lugt, F.M. M\"uller, Phys. Rev. B
{\bf 32},
 2319 (1985).
\bibitem{Geertsma} W. Geertsma, J. Phys. C {\bf 18}, 2461
(1985).
\bibitem{Sab} J. Fortner, M.-L. Saboungi, and J.E.
 Enderby, Phys. Rev. Lett. {\bf 74}, 1415 (1995).
\end{thebibliography}
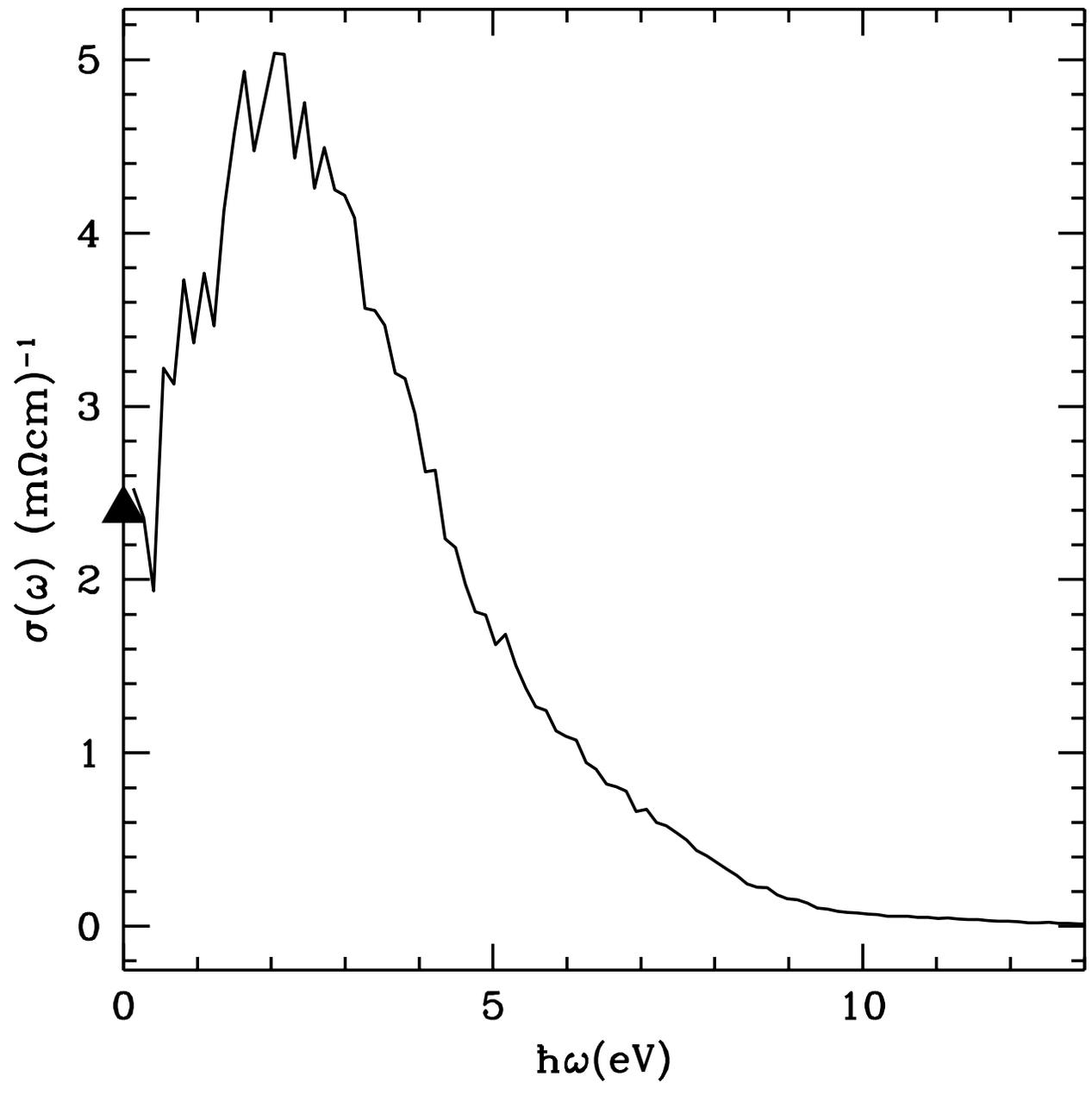
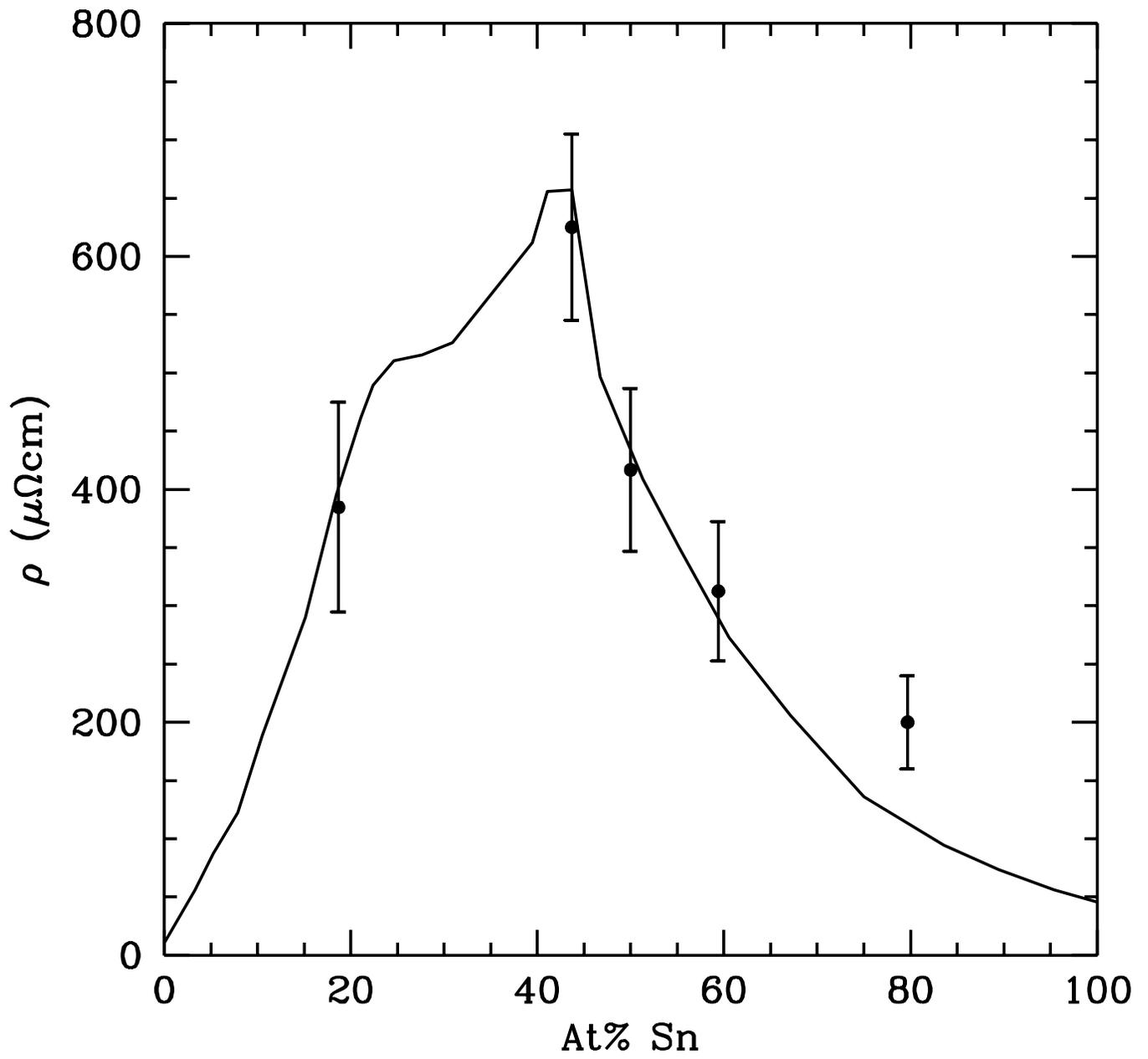
\end{document}